\documentclass[10pt]{article}

\usepackage{amsmath}
\usepackage{amssymb}

\usepackage{graphicx}

\usepackage{cite}

\usepackage{color} 

\usepackage[labelfont=bf,labelsep=period,justification=raggedright]{caption}

\makeatletter
\renewcommand{\@biblabel}[1]{\quad#1.}
\makeatother

\date{}

\begin{document}

\begin{flushleft}
{\Large
\textbf{Individualization as driving force of clustering phenomena in humans}
}
\\
Michael M\"as$^{1,\ast}$, 
Andreas Flache$^{1}$, 
Dirk Helbing$^{2,3,4}$
\\
\bf{1} Department of Sociology/ICS, University of Groningen, Groningen, The Netherlands
\\
\bf{2} Chair of Sociology, in particular of Modeling and Simulation, ETH Zurich, Zurich, Switzerland
\\
\bf{3} Santa Fe Institute, Santa Fe, New Mexico, United States of America
\\
\bf{4} Collegium Budapest - Institute for Advanced Study, Budapest, Hungary
\\
$\ast$ E-mail: m.maes@rug.nl
\end{flushleft}

\section*{Abstract}
One of the most intriguing dynamics in biological systems is the emergence of clustering, the self-organization into separated agglomerations of individuals. Several theories have been developed to explain clustering in, for instance, multi-cellular organisms, ant colonies, bee hives, flocks of birds, schools of fish, and animal herds. 
A persistent puzzle, however, is clustering of opinions in human populations. The puzzle is particularly pressing if opinions vary continuously, such as the degree to which citizens are in favor of or against a vaccination program. Existing opinion formation models suggest that ``monoculture'' is unavoidable in the long run, unless subsets of the population are perfectly separated from each other. Yet, social diversity is a robust empirical phenomenon, although perfect separation is hardly possible in an increasingly connected world. 
Considering randomness did not overcome the theoretical shortcomings so far. Small perturbations of individual opinions trigger social influence cascades that inevitably lead to monoculture, while larger noise disrupts opinion clusters and results in rampant individualism without any social structure.
\par
Our solution of the puzzle builds on recent empirical research, combining the integrative tendencies of social influence with the disintegrative effects of individualization. A key element of the new computational model is an adaptive kind of noise. We conduct simulation experiments to demonstrate that with this kind of noise, a third phase besides individualism and monoculture becomes possible, characterized by the formation of metastable clusters with diversity \textit{between} and consensus \textit{within} clusters. When clusters are small, individualization tendencies are too weak to prohibit a fusion of clusters. When clusters grow too large, however, individualization increases in strength, which promotes their splitting. 
\par
In sum, the new model can explain cultural clustering in human societies. Strikingly, our model predictions are not only robust to ``noise''---randomness is actually the central mechanism that sustains pluralism and clustering.


\section*{Introduction}
Many biological systems exhibit collective patterns, which emerge through simple interactions of large numbers of individuals. A typical example are agglomeration phenomena. Such clustering dynamics have been found in systems as different as bacterial colonies \cite{BenJacob}, gregarious animals like cockroaches \cite{Theraulaz}, fish schools \cite{Fish}, flocks of birds \cite{Ballerini}, and animal groups \cite{Couzin}. Similar phenomena are observed in ecosystems \cite{Levin} and human populations, as the formation of urban agglomerations demonstrates \cite{Stanley,Batty1}.
\par
Recently, the clustering of social interactions in human populations \cite{Watts1998,Barabasi1999,Palla2007} has been extensively studied in networks of email communication\cite{LibenNovell2008}, phone calls\cite{Palla2007}, scientific collaboration \cite{Newman2004} and sexual contacts\cite{Liljeros2001}. It is much less understood, however, how the pattern of \textit{opinion clustering} that characterizes modern societies comes about. Empirical studies suggest that in modern societies opinions differ globally \cite{Fiorina2008}, while they cluster locally within geographical regions \cite{Glaeser2006}, socio-demographic groups \cite{Mark2003}, or internet communities \cite{Lazer2009}. The lack of a theoretical understanding of opinion clustering is pressing, since both, local consensus and global diversity are precarious. On the one hand, cultural diversity may get lost in a world where people are increasingly exposed to influences from mass media, Internet communication, interregional migration, and mass tourism, which may promote a universal monoculture \cite{Friedman2005,Greig2002}, as the extinction of languages suggests \cite{Sutherland2003}. On the other hand, increasing individualization threatens to disintegrate the social structures in which individuals are embedded, with the possible consequence of the loss of societal consensus \cite{Durkheim1997,Beck1994}. This is illustrated by the decline of the social capital binding individuals into local communities \cite{McPerson2006}.
\par
Early formal models of social influence imply that monoculture is unavoidable, unless a subset of the population is perfectly cut off from outside influences \cite{Abelson1964}. Social isolation, however, appears questionable as explanation of pluralism. In modern societies, distances in social networks are quite short on the whole, and only relatively few random links are required to dramatically reduce network distance \cite{Watts1998}. 
\par
Aiming to explain pluralism, researchers have incorporated the empirically well-supported observation of ``homophily'', i.e. the tendency of ``birds of a feather to flock together'' \cite{Pherson2001,Aral2009}, into formal models of social influence \cite{Nowak1990}. These models typically assume ``bounded confidence'' (BC) in the sense that only those individuals interact, whose opinions do not differ more than a given threshold level \cite{Hegselmann2002,Deffuant2005}. As Fig. 1A illustrates, BC generates opinion clustering, a result that generalizes to model variants with categorical rather than continuous opinions \cite{Nowak1990,Axelrod1997}. However, clustering in the BC-model is sensitive to ``interaction noise'': A small random chance that agents may interact even when their opinions are not similar, causes monoculture again (see Fig. 1B). 
\par
To avoid this convergence of opinions, it was suggested that individuals would separate themselves from negatively evaluated others \cite{Mark2003,Macy2003}. However, recent empirical results do not support such ``negative influence''  \cite{Krizan2007}. Scientists also tried to avoid convergence by ``opinion noise'', i.e. random influences, which lead to arbitrary opinion changes with a small probability. Assuming uniformly distributed opinion noise \cite{Pineda2009} leads to sudden, large, and unmotivated opinion changes of individuals, while theories of social integration \cite{Durkheim1997,Beck1994,Hornsey2004,Vignoles2000} and empirical studies of individualization \cite{Snyder1980,Imhoff2009} show a tendency of incremental opinion changes rather than arbitrary opinion jumps. Incremental opinion changes, however, tend to promote monoculture, even in models with categorical rather than continuous opinions \cite{Klemm2003}. Figure 1 demonstrates that adding a ``white noise'' term ($N(0,\theta)$) to an agent's current opinion in the BC model fails to explain opinion clustering. Weak opinion noise ($\theta=5$) triggers convergence cascades that inevitably end in monoculture. Stronger noise restores opinion diversity, but not pluralism. Instead, diversity is based on frequent individual deviations from a predominant opinion cluster (for $\theta=18$). However, additional clusters can not form and persist, because opinion noise needs to be strong to separate enough agents from the majority cluster---so strong that randomly emerging smaller clusters cannot stabilize. 
\par
In conclusion, the formation of persistent opinion clusters is such a difficult puzzle that all attempts to explain them had to make assumptions that are difficult to justify by empirical evidence. The solution proposed in the following, in contrast, builds on sociological and psychological research.  The key innovation is to integrate another decisive feature into the model, namely the ``strive for uniqueness'' \cite{Snyder1980,Imhoff2009}. While individuals are influenced by their social environment, they also show a desire to increase the uniqueness when too many other members of society hold similar opinions. This  suggests that noise in individual opinion formation is adaptive, creating a dynamic interplay of the integrating and disintegrating forces highlighted by Durkheim's classic theory of social integration \cite{Durkheim1997}. Durkheim argued that integrating forces bind individuals to society, motivating them to conform and adopt values and norms that are similar to those of others. But he also saw societal integration as being threatened by disintegrating forces that foster individualization and drive actors to differentiate from one another \cite{Beck1994,Hornsey2004,Vignoles2000}.  The continuous ``Durkheimian opinion dynamics model'' proposed in the following can explain pluralism, although it incorporates all the features that have previously been found to \textit{undermine} clustering: (1) a fully connected influence network, (2) absence of bounded confidence, (3) no negative influence, and (4) white opinion noise. From a methodological viewpoint, our model builds on concepts from statistical physics, namely the phenomenon of ``nucleation'' \cite{Stanley1971}, illustrated by the formation of water droplets in supersaturated vapor. However, by assuming adaptive noise, we move beyond conventional nucleation models. 
\par
Computational experiments reveal that our model generates pluralism as an intermediate phase between monoculture and individualism. When the integrating forces are too strong, the model dynamics inevitably implies monoculture, even when the individual opinions are initially distributed at random. When the disintegrating forces prevail, the result is what Durkheim called ``anomie'', a state of extreme individualism without a social structure, even if there is perfect consensus in the beginning. Interestingly, there is no sharp transition between these two phases, when the relative strength of both forces is changed. Instead, we observe an additional, intermediate regime, where opinion clustering occurs, which is independent of the initial condition. In this regime, adaptive noise entails robust pluralism that is stabilized by the adaptiveness of cluster size. When clusters are small, individualization tendencies are too weak to prohibit a fusion of clusters. However, when clusters grow large, individualization increases in strength, which triggers a splitting into smaller clusters (``fission''). In this way, our model solves the cluster formation problem of earlier models. While in BC models, white noise causes either monoculture or fragmentation (Fig. 1C), in the Durkheimian opinion dynamics model proposed here, it \textit{enables} clustering. Therefore, rather than \textit{endangering} cluster formation, noise supports it. In the following, we describe the model and identify conditions under which pluralism can flourish.

\section*{Model}
The model has been elaborated as an agent-based model \cite{Bonabeau2002} addressing the opinion dynamics of interacting individuals. The simulated population consists of $N$ agents $i$, representing individuals, each characterized by an opinion $o_i(t)$ at time $t$. The numerical value for the opinion varies between a given minimum and maximum value on a metric scale. 
We use the term ``opinion'' here, for consistency with the literature on social influence models. However, $o_i$ may also reflect behaviors, beliefs, norms, customs or any other cardinal cultural attribute that individuals consider relevant and that is changed by social influence. The dynamics is modeled as a sequence of events. Every time $t'=k/N$ (with $k\in \{1,...,N\}$), the computer randomly picks an agent $i$ and changes the opinion $o_i$ by the amount
\begin{equation}
 \Delta o_i = \frac{\displaystyle \sum_{j=1\atop j\ne i}^N \big(o_j - o_i\big) w_{ij}}{\displaystyle \sum_{j=1\atop j\ne i}^N w_{ij}} + \xi_i \, .
 \label{eins}
\end{equation}
The first term on the rhs of Eq. [\ref{eins}] models the integrating forces of Durkheim's theory. Technically, agents tend to adopt the weighted average of the opinions $o_j$ of all other members $j$ of the population. Implementing homophily, the social influence $w_{ij}$ that agent $j$ has on agent $i$ is the stronger, the smaller their opinion distance $d_{ij} = |o_j - o_i|$ is. Formally, we assume
\begin{equation} 
 w_{ij} = \mbox{e}^{- d_{ij}/A} = \mbox{e}^{- |o_j - o_i|/A}\, . 			
 \label{zwei}
\end{equation}
The parameter $A$ represents the range of social influence of agents. For small positive values of $A$, agents are very confident in their current opinion and are mainly influenced by individuals who hold very similar opinions, while markedly distinct opinions have little impact. The higher $A$ is, however, the more are agents influenced by individuals with considerably different opinions and the stronger are the integrating forces in our Durkheimian theory.
\par
The \textit{dis}integrating forces on the opinion of agent $i$ are modeled by a noise term $\xi_i$. Specifically, the computer adds a normally distributed random value $\xi_i$ (``white noise'') to the first term on the rhs of Eq. [\ref{eins}]. While we assume that the mean value of the random variable $\xi_i$ is zero, the standard deviation has been specified as
\begin{equation}
 \theta_{it} = s\sum_{j=1}^N \mbox{e}^{-d_{ij}} \, . 
\label{drei}
\end{equation}

The larger the standard deviation, the stronger are the individualization tendencies of an agent. Following Durkheim's theory, equation [\ref{drei}] implements the assumption that an agent's strive for individualization is weak, if there are only a few others with similar opinions.  Under such conditions, there is no need to increase distinctiveness. However, if many others hold a similar opinion, then individuals are more motivated to differ from others.
\par
By including the focal agent $i$ in the sum of Eq. [\ref{drei}], we assume that there is always some opinion noise, even when agent $i$ holds a perfectly unique opinion. These fluctuations may have a variety of reasons, such as misjudgements, trial-and-error behavior, or the influence of exogenous factors on the individual opinion. 
\par
We use the parameter $s$ to vary the strength of the disintegrating forces in society. The higher the value of $s$, the higher is the standard deviation of the distribution, from which $\xi_i$ is drawn, and the stronger are the disintegrating forces. Finally, to keep the opinions of the agents within the bounds of the opinion scale, we set the value of $\xi_i$ to zero, if the bounds of the opinion space would be left otherwise.

\section*{Results}
We have studied the Durkheimian opinion dynamics model with extensive computer simulations, focussing ourselves on relatively small populations ($N=100$), because in this case it is reasonable to assume that all members may interact with each other. For bigger groups one would have to take into account the topology of the social interaction network as well. Such networks would most likely consist of segregated components (``communities''), which are not or only loosely connected with each other \cite{Palla2007,LibenNovell2008,Newman2004,Liljeros2001}. Because of the weak or missing connections \textit{between} communities, it would not be so surprising if each community developed its own, shared opinion. In small, completely connected populations, however, the occurrence of diverse opinions is puzzling, as it cannot result from a lack of contacts between agents.
\par
To illustrate the model dynamics, Fig. 2 shows three typical simulation runs for different strengths $s$ of disintegrating forces, while the strength $A=2$ of the integrating force is kept constant. In each run, all agents start with an opinion in the middle of the opinion scale ($o_i=0$), i.e. conformity. This is an initial condition for which the classical BC-model does not produce diversity. Fig. 2A shows typical opinion trajectories for a population, in which the integrating forces are much stronger than the disintegrating forces. Consequently, the population develops collective consensus, i.e. the variation of opinions remains small, even though not all agents hold exactly the same opinion. Triggered by the random influences $\xi_i$, the average opinion performs a characteristic random walk. 
\par
When the disintegrating force prevails, the pattern is strikingly different. Fig. 2B shows that for large noise strengths $s$, the initial consensus breaks up quickly, and the agents' opinions are soon scattered across the entire opinion space.
\par
Simulation scenarios A and B are characteristic for what Durkheim referred to as states of social cohesion and of anomie. Interestingly, however, pluralism arises as a third state in which several opinion clusters form and coexist. Fig. 2C shows a typical simulation run, where the adaptive noise maintains pluralism depite the antagonistic impacts of integrating and disintegrating forces---in fact \textit{because} of this. In the related region of the parameter space, disintegrating forces prevent global consensus, but the integrating forces are strong enough to prevent the population from extreme individualization. This is in pronounced contrast to what we found for the BC-model with strong noise (Fig. 1C). Instead, we obtain a number of coexisting, metastable clusters of a characteristic, parameter-dependent size. Each cluster consists of a relatively small number of agents, which keeps the disintegrating forces in the cluster weak and allows clusters to persist. (Remember that the tendency of individualization according to Eq. [\ref{drei}] increases, when many individuals hold similar opinions.) However, due to opinion drift, distinct clusters may eventually merge. When this happens, the emergent cluster becomes unstable and will eventually split up into smaller clusters, because disintegrating forces increase in strength as a cluster grows. 
\par
Strikingly, the state of diversity, in which several opinion clusters can coexist, is not restricted to a narrow set of conditions under which integrating and disintegrating forces are balanced exactly. Fig. 3 demonstrates that opinion clusters exist in a significant area of the parameter space, i.e. the clustering state establishes another phase, which is to be distinguished from monoculture and from anomie. 
\par
To generate Fig. 3, we conducted a simulation experiment in which we varied the influence range $A$ and the strength $s$ of the disintegrating force. For each parameter combination, we ran 100 replications and measured the average number of clusters that were present after 250,000 iterations. To count the number of clusters in a population, we ordered the $N$ agents according to their opinion. A cluster was defined as a set of agents in adjacent positions such that each set member was separated from the adjacent set members by a maximum of 5 scale points (= opinion range/$N$). Fig. 3 shows that, for large social influence ranges $A$ and small noise strengths $s$, the average number of clusters is below 1.5, reflecting monoculture in the population. In the other extreme, i.e. for a small influence range $A$ and large noise strengths $s$, the resulting distribution contains more than 31 clusters, a number of clusters that cannot be distinguished from purely random distributions. Following Durkheim, we have classified such cases as anomie, i.e. as the state of extreme individualism. Between these two phases, there are numerous parameter combinations, for which the number of clusters is higher than 1.5 and clearly smaller than in the anomie phase. This constitutes the clustering phase. Fig. 3 also shows that, for each parameter combination, there is a small variance in the number of clusters, which is due to a statistical equilibrium of occasional fusion and fission processes of opinion clusters (see Fig. 2C).
\par 
The same results were found, when starting the computer simulations with a uniform opinion distribution. Additional statistical tests were performed to make sure that the existence of clusters in our model indeed indicates pluralism and not fragmentation, a state in which a population consists of one big cluster and a number of isolated agents (see Fig. 4). While the Durkheimian opinion dynamics model is consistent with cluster formation (see Fig. 4A), the noisy BC model rather shows random fragmentation (see Fig. 4B).


\section*{Discussion}
The phenomenon of self-organized clustering phenomena in biological and social systems is widespread and important. With the advent of mathematical and computer models for such phenomena, there has been an increasing interest to study them also in human populations. The work presented here focuses on resolving the long-standing puzzle of opinion clustering.
\par
The emergence and persistence of pluralism is a striking phenomenon in a world in which social networks are highly connected and social influence is an ever present force that reduces differences between those who interact. We have developed a formal theory of social influence that, besides anomie and monoculture, shows a third, pluralistic phase characterized by opinion clustering. It occurs, when all individuals interact with each other and noise prevents the convergence to a single opinion,  despite homophily.
\par
Our model does not assume negative influence, and it behaves markedly different from bounded confidence models, in which white opinion noise produces fragmentation rather than clustering. It would be natural to generalize the model in a way that also considers the structure of real social networks. This basically requires one to replace the values $w_{ij}$ by  $w_{ij}a_{ij}$, where $a_{ij}$ are the entries of the adjacency matrix (i.e. $a_{ij} = 1$, if individuals $i$ and $j$ interact, otherwise $a_{ij}=0$). In such a case, resulting opinion clusters are expected to have a broad range of different sizes, similar to what is observed for the sizes of social groups.
\par
Our model highlights the functional role that ``noise'' (randomness, fluctuations, or other sources of variability) plays for the organization of social systems. It furthermore shows that the combination of two mechanisms (deterministic integrating forces and stochastic disintegrating forces) can give rise to new phenomena. We also believe that our results are meaningful for the analysis of the social integration of our societies. Both classical \cite{Durkheim1997} and contemporary \cite{Beck1994} social thinkers argue that, in modern and globalized societies, individuals are increasingly exposed to disintegrating forces that detach them from traditional social structures. In other words, the social forces that motivate individuals to follow societal norms may lose their power to limit individual variation. Durkheim feared that this will atomize societies \cite{Durkheim1997}. That is, he thought the tendency towards individualization would turn societies ``anomic'' as they modernize: Durkheim felt that extreme individualization in modern societies would obstruct the social structures that traditionally provided social support and guidance to individuals. 
\par
Today, modern societies are highly diverse, but at the same time they are far from a state of anomie as foreseen by Durkheim. Our model offers an explanation why and how this is possible: Besides monoculture and anomie, there is a third, pluralistic clustering phase, in which individualization prevents overall consensus, but at the same time, social influence can still prevent extreme individualism. The interplay between integrating and disintegrating forces leads to a plurality of opinions, while metastable subgroups occur, within which individuals find a local consensus. Individuals may identify with such subgroups and develop long-lasting social relationships with similar others. Therefore, they are not isolated and not without support or guidance, in contrast to the state of anomie that Durkheim was worried about.
\par
We have seen, however, that pluralism and cultural diversity require an approximate balance between integrating and disintegrating forces. If this balance is disturbed, societies may drift towards anomie or monoculture. It is, therefore, interesting to ask how the current tendency of globalization will influence society and cultural dynamics. The Internet, interregional migration, and global tourism, for example, make it easy to get in contact with members of distant and different cultures. Previous models \cite{Axelrod1997, Greig2002} suggest that this could affect cultural diversity in favor of a monoculture. However, if the individual strive for uniqueness is sufficiently strong, formation of diverse groups (a large variety of international social communities) should be able to persist even in a globalizing world. In view of the alternative futures, characterized by monoculture or pluralism, further theoretical, empirical, and experimental research should be performed to expand our knowledge of the mechanisms that will determine the future of pluralistic societies.




\section*{Figures}

\begin{figure}[!ht]
\begin{center}
\includegraphics[width=17.35cm]{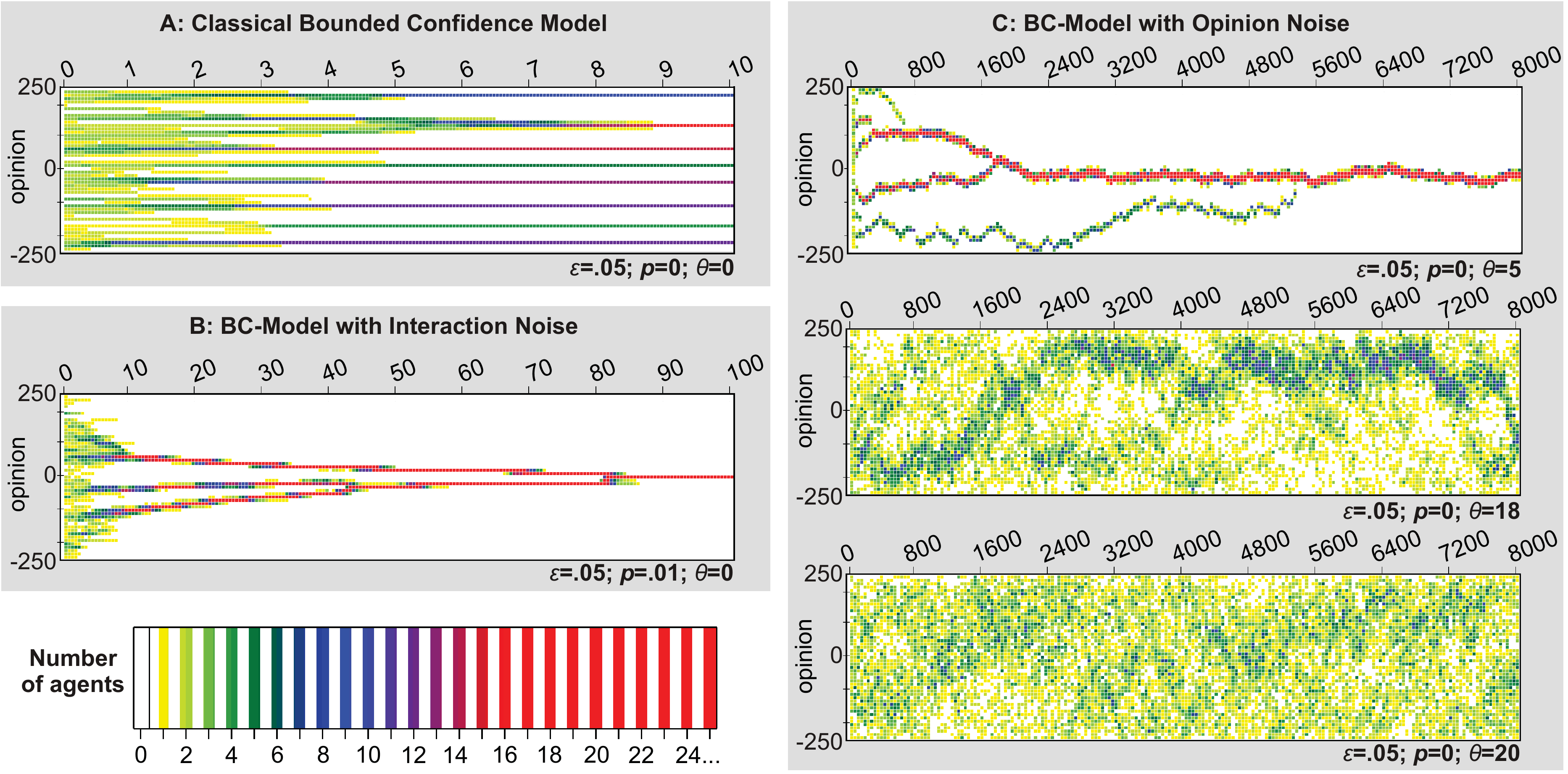}
\end{center}
\caption{ {\bf Opinion dynamics produced by the bounded confidence (BC) model \cite{Hegselmann2002} with and without noise}. Populations consist of 100 agents. Opinions vary between -250 and 250. Initial opinions are uniformly distributed.
For visualization, the opinion scale is divided into 50 bins of equal size. Color coding indicates the relative frequency of agents in each bin. (A) Dynamics of the \textit{BC-model without noise} \cite{Hegselmann2002} over 10 iterations. At each simulation event, one agent's opinion is replaced by the average opinion of those other agents who hold opinions $o_{j}$ within the focal agent's confidence interval ($o_{i}-\epsilon \le o_{j} \le o_{i}+\epsilon$). For $\epsilon =0.05$, one finds several  homogeneous clusters, which stabilize when the distance between all clusters exceeds the confidence threshold $\epsilon$.
(B) Computer simulation of the same BC-model, but considering \textit{interaction noise}. Agents that would otherwise not have been influential, now influence the focal agent's opinion with a probability of $p=0.01$. This small noise is sufficient to eventually generate monoculture.
(C) Simulation of the BC-model with opinion noise. After each opinion update, a normally distributed random value drawn from $N(0,\theta)$ is added to the opinion. Under weak opinion noise ($\theta=5$), one cluster is formed, which carries out a random walk on the opinion scale. 
When the opinion noise is significantly increased ($\theta=18$), there is still one big cluster, but many separated agents exist as well (cf. Fig. 4). With even stronger opinion noise ($\theta=20$), the opinion distribution becomes completely random.}
\label{Fig1}
\end{figure}

\begin{figure}[!ht]
\begin{center}
\includegraphics[width=17.35cm]{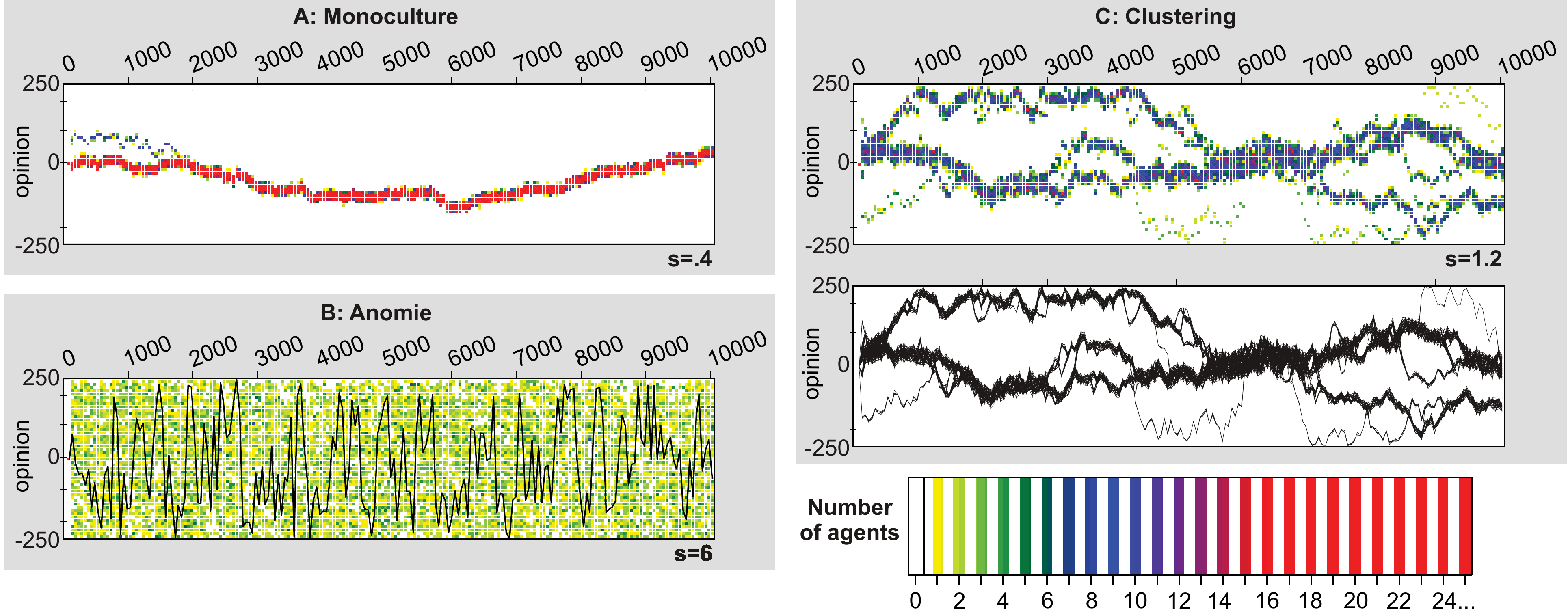}
\end{center}
	\caption{{\bf Opinion trajectories of three representative simulation runs with 100 agents generated by the Durkheimian model.} In all three runs, the opinions are restricted to values between -250 and 250, and all agents hold the same opinion initially ($o_{i}(0)=0$ for all $i$). In all runs, we assume the same social influence range $A=2$, but vary the strength $s$ of the disintegrating force. (A) Monoculture, resulting in the case of a weak disintegrating force ($s=0.4$). Agents do not hold perfectly identical opinions, but the variance is low. We studied dynamics over 10.000 iterations.
	(B) Anomie (i.e. extreme individualism), generated by a very strong disintegrating force ($s=6$). Agents spread over the complete opinion scale. The black line represents the time-dependent opinion of a single, randomly picked agent, showing significant opinion changes over time, which is in contrast to the collective opinion formation dynamics found in the monocultural and pluralistic cases (A) and (B). 	
	(C) For a moderate disintegrating force ($s=1.2$), the population quickly disintegrates into clusters. As long as these clusters are small, they are metastable. However, clusters perform random walks and can merge (e.g. around iteration 5500). As the disintegrating force grows with the size of a cluster, big clusters eventually split up into subclusters (e.g. around iteration 7000). The additional graph, in which each agent's opinion trajectory is represented by a solid black line, is an alternative visualization of the simulation run with $s=1.2$. It shows that the composition of clusters persists over long time periods. 
}
\label{Fig2}
\end{figure}

\begin{figure}[!ht]
\begin{center}
\includegraphics[width=8.3cm]{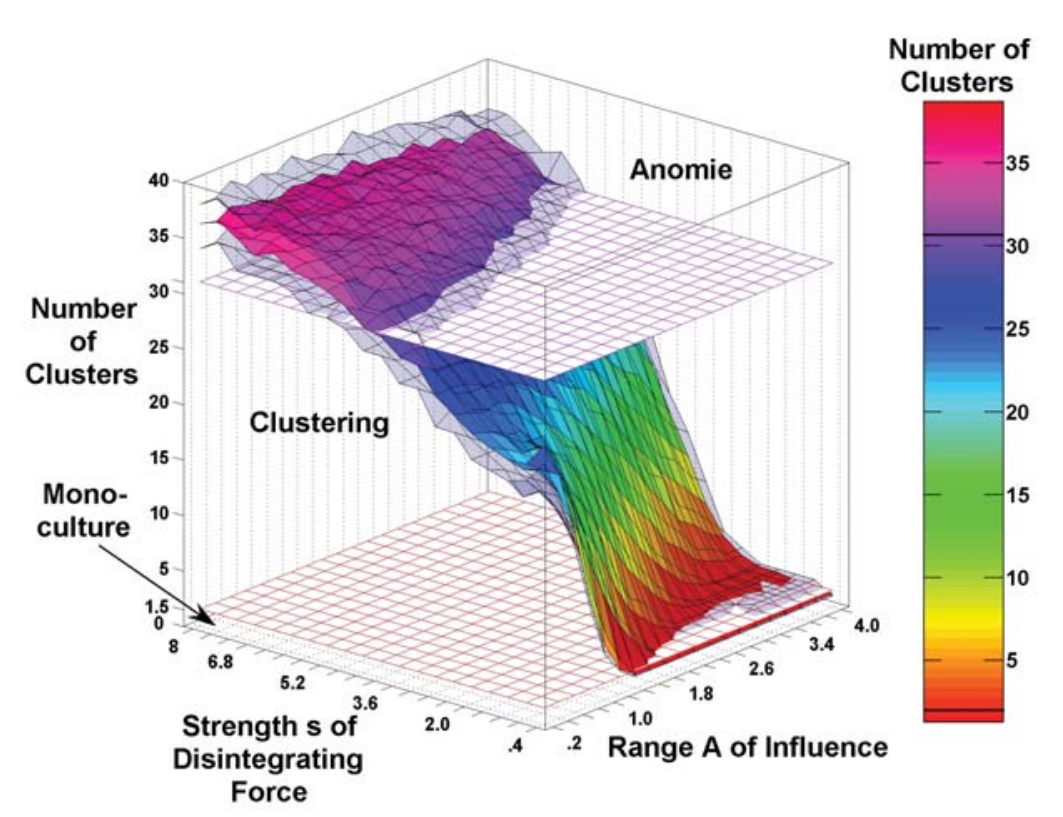}
\end{center}
	\caption{{\bf Conditions of clustering, monoculture and Anomie.} The figure shows the dependence of the average number of clusters in the Durkheimian model on the strength $s$ of the disintegrating force and the range $A$ of social influence.	To generate it, we conducted computer simulations with $N=100$ agents, starting with initial consensus ($o_{i}(0)=0$ for all $i$). We restricted opinions to values between -250 and 250. We varied the strength $s$ of the disintegrating force between $s=0.4$ and $s=8$ in steps of 0.4. $A$ varied between $A=0.2$ and $A=4$ in steps of 0.2. For each parameter combination, we conducted 100 independent replications and assessed the average number of clusters formed after 250,000 iterations (see $z$-axis and the color scale). The two transparent (gray) surfaces depict the inter-quartile range, which indicates a small variance in the number of clusters (and also typical cluster sizes) for each parameter combination. The horizontal grids indicate the borders of the three phases, as defined by us. An average cluster size below 1.5 indicates monoculture. Values between 1.5 and 31 reflect clustering. Finally, values above 31 correspond to opinion distributions that cannot be distinguished from random ones and represent a state of anomie. 	
}
\label{Fig3}
\end{figure}

\begin{figure}[!ht]
\begin{center}
\includegraphics[width=17.35cm]{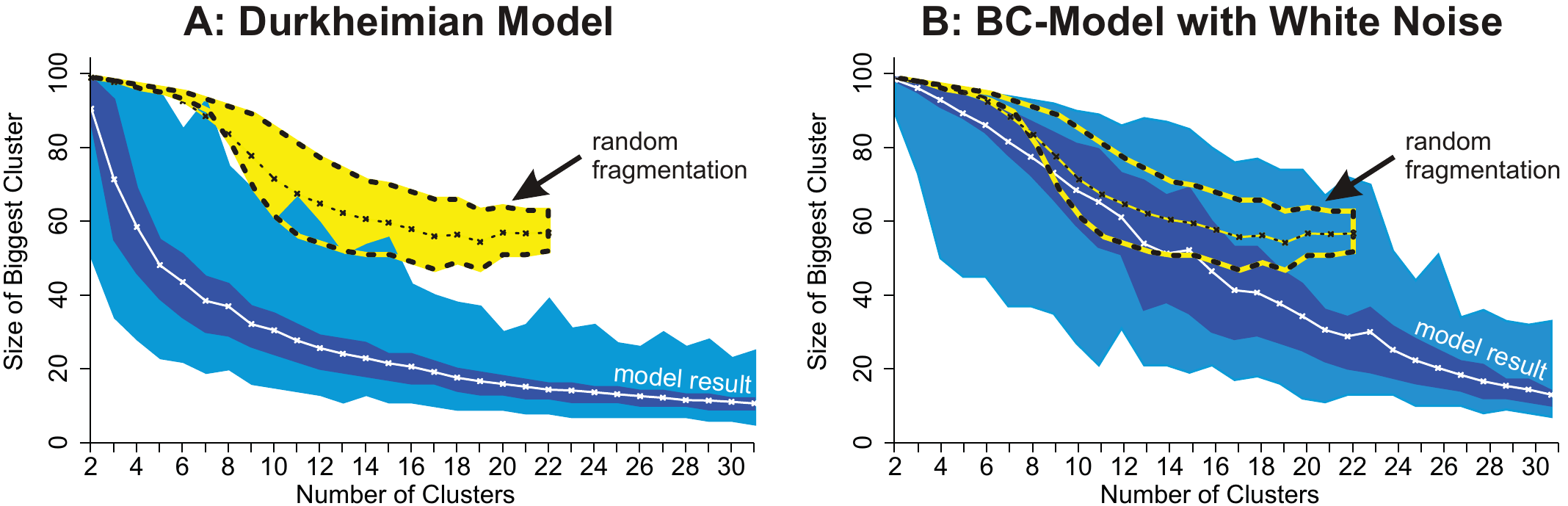}
\end{center}
	\caption{ {\bf Comparison of the (A) Durkheimian model and (B) the noisy BC-model.} Figures show the size of the biggest cluster over the number of clusters (in all simulation runs that resulted in more than one and less than 32 clusters). Fig. 4A is based on the simulation experiment underlying Figure 3. Figure 4B was generated by an experiment with the BC-model \cite{Hegselmann2002}, in which we varied the bounded-confidence level $\epsilon$ between 0.01 and 0.15 in steps of 0.02 and the noise level $\theta$ between 5 and 50 in steps of 5. We conducted 100 replications per parameter combination and measured the number of clusters and the size of the biggest cluster after 250,000 iterations. 
White solid lines represent the average size of the biggest cluster. The dark blue area shows the respective interquartile range and the light blue area the complete value range. For comparison, we generated randomly fragmented opinion distributions of $N=100$ individuals as follows: $n$ agents were assigned to hold a random opinion drawn from a normal distribution ($N(0,50)$). The remaining $N-n$ agents were assumed to hold opinion $o_i=0$, thereby forming one big cluster. We varied the value of $n$ between 0 and 100 in steps of 1 and generated 1000 distributions per condition. The average size of the biggest cluster of the resulting distributions is shown by the thin yellow-black line. (The curve stops at 22, since this is the highest number of clusters generated.) The bold yellow-black lines represent the related interquartile range. 
We find that the value range of the Durkheimian model (blue area) hardly overlaps with the interquartile range of the fragmented distributions (yellow area). This demonstrates that the Durkheimian model shows clustering rather than fragmentation. In contrast, Fig. 4B illustrates that the distributions of the noisy BC-model and the results for random fragmentation overlap.
}
\label{Fig4}
\end{figure}


\end{document}